\documentclass{llncs}
\usepackage{graphicx}
\usepackage{color}
\usepackage{enumitem}
\usepackage{hyperref}
\usepackage{dirtytalk}
\usepackage[utf8]{inputenc}
\hypersetup{
    colorlinks=true,
    linkcolor=blue,
    filecolor=magenta,      
    urlcolor=cyan,
}
\usepackage[textsize=tiny]{todonotes} 

\definecolor{mygreen}{HTML}{00CC00}

\begin{document}

\title{LITMUS: An Open Extensible Framework for Benchmarking RDF Data Management Solutions}

\author{Harsh Thakkar\inst{1}, Mohnish Dubey\inst{1}, Gezim Sejdiu\inst{1}, Axel-Cyrille Ngonga Ngomo\inst{2}, Jeremy Debattista\inst{1,3}, Christoph Lange\inst{1,3}, Jens Lehmann\inst{1,3}, S\"{o}ren Auer\inst{1,3}, Maria-Esther Vidal\inst{1,3}}
\institute{
	University of Bonn; \\
	\email{\{thakkar,dubey,sejdiu,debattis,langec,jens.lehmann,auer,vidal\}@cs.uni-bonn.de}
	\and 
	University of Leipzig; \\
	\email{ngonga@informatik.uni-leipzig.de}
	\and 
	Fraunhofer IAIS;
\vspace{-10pt}
	}
\maketitle

\begin{abstract}
Developments in the context of Open, Big, and Linked Data have led to an enormous growth of structured data
on the Web. 
To keep up with the pace of efficient consumption and management of the data at this rate, many data management solutions have been developed for specific tasks and applications.
We present LITMUS, a framework for benchmarking data management solutions.
LITMUS goes beyond classical storage benchmarking frameworks by allowing for analysing the performance of frameworks across query languages.
In this position paper we present the conceptual architecture of LITMUS as well as the considerations that led to this architecture.
\end{abstract}

\section{Introduction}\label{sec:Introduction}
     
Vast amounts of structured (following Linked Data principles) and un/semi-structured data is constantly being made available on the Web, often in an open manner\footnote{With \emph{open} we follow the Open Data Definition~\url{http://opendefinition.org}}, and within organisations.
This rapid growth of data, available across organisations, has affected the data management layer of modern applications. 
Consequently,
organisations are increasingly facing the need to find data management tools suited for the specific tasks at the core of their information management. 
Choosing the best data management tool is, however, challenging due to the limited comparability and compatibility of existing evaluation results and benchmarks. 
With regard to the limited domain expertise of the end user, 
the need for standardised frameworks to benchmark and analyse the existing diverse data management platforms is consequently of paramount importance.

Despite the growing interest and use in both research and the industry communities, currently 
the creators of benchmarks for Data Management Solutions (DMS)~\cite{alucc2014diversified,Bizer2009TheBS} do not offer a common 
suite for performing cross-domain benchmarks (i.e. one-to-one comparison of RDF, Graph, Wide-column, Relational, etc stores). In addition, there is no significant baseline to compare these cross-domain DMSs one against the other. 
Moreover, reproducing benchmarks is a non-trivial problem owing to reasons such as non-standardised setup configurations, lack of publicly available resources (such as scripts, libraries, packages, etc.) and lack of transparent evaluation policies. Results in areas such as named entity recognition and linking~\cite{gerbil} as well as question answering~\cite{bioasq,qald} have, however, shown that the provision of standardised interfaces and measures can contribute to the improvement of the performance of software solutions. 
    
In this position paper we present the concept behind LITMUS, an open extensible approach for benchmarking a wide variety of DMS for storing RDF. 
LITMUS aims to provide support to organisations aspiring to use Linked Data management technologies in a wide spectrum of applications and magnitudes. 
LITMUS will provide a realistic performance evaluation platform covering a plethora of heterogeneous technologies (see Section~\ref{litmus_framework}) for storage and query benchmarking. 
To put the reader into the context of this work, and to highlight the objectives of LITMUS, we present the following \textbf{user scenario}: 
    
    {\small \say{The WDAqua research project\footnote{WDAqua ITN -- \url{http://wdaqua.eu}} aims towards building a data-driven question answering platform by using Web data, available in various formats, e.g., RDF, CSV, SQL, or XML. Harsh, a researcher within the project,  is responsible for ensuring efficient data management (storage and retrieval) for this project. 
    There is a large number of DMSs, each deliberately tailored to handling specific formats of data and queries, which need to be benchmarked to select the best solution for the project's needs. 
    However, benchmarking of DMSs is non-trivial: it takes large amounts of human effort in designing, administering, evaluating, and analysing the diverse systems involved. 
    Additionally, for the research project, a large set of 
    factors, e.g., query typology, indexing speed, index size, query response time, and dataset size, need to be considered to ensure reproducibility and generality of the observed experimental results.
    \textit{Harsh, wants to automate the whole benchmarking process, allowing easy integration, evaluation on custom stress loads, and fast analysis of the evaluation results}. He would also \textit{expect the framework to be flexible to integrate new DMSs to the plethora of existing systems and benchmark them against a baseline}.
    Thus, Harsh's research question is: \textit{Can a computational framework provide the required support for identifying the independent factors of his experiments, and for analysing and interpreting of the experimental results?}.
    The answer to this research question is \textit{yes}, and the computational framework is LITMUS, an open extensible platform for benchmarking cross-domain DMSs.
    LITMUS will not only satisfy Harsh’s need for automating the tedious benchmarking process, but will also offer: \textit{\textbf{(1)}} an efficient way for replicating existing benchmarks (e.g., BSBM~\cite{Bizer2009TheBS} or WAT-DIV~\cite{alucc2014diversified}); \textit{\textbf{(2)}} a wide set of performance evaluation measures/indicators tailored specifically for the DMS being evaluated; and \textit{\textbf{(3)}} the comparison and visualisation of the performance of benchmarked DMSs on various intrinsic factors via custom charts, graphs and tabular data.}}

The remainder of this article is organised in the following sections: 
(\ref{relwork}) \textit{Related work} on benchmarking efforts, and their shortcomings, 
(\ref{Objectives}) \textit{Objectives, Challenges and Outcomes}, which shed light on the focus of LITMUS,
(\ref{litmus_framework}) \textit{Framework}, describing the components the LITMUS components, and
(\ref{conclusion}) \textit{Conclusions}, summarizing the article.

\section{Related work}\label{relwork}

Benchmarking is widely used for evaluating data stores. 
Benchmarks exist for a variety of levels of abstraction from simple data models to graphs and triple stores, to entire enterprise information systems.
We describe the current state of the art in benchmarking, in particular benchmarks for (a) relational databases, (b) graph databases, (c) RDF stores, (d) key-value stores, (e) wide-column stores, and (f) cross-domain benchmarking efforts.
We identify shortcomings and limitations of existing systems, in order to determine the gaps that LITMUS needs to take into consideration.
In addition to surveying existing work, we intend to focus mainly on the purpose and scope of the benchmarks.
    
    In \textit{Relational} DMSs, the benchmarks of the Transaction Processing Performance Council (\textbf{TPC})~\cite{Nambiar2011} are well established.
    TPC uses discrete metrics for measuring the performance of the relational DMS.
    The online transaction processing benchmarks \textbf{TPC-C} and \textbf{TPC-E} use a transactions per minute metric.
    The analytics \textbf{TPC-H} and decision support \textbf{TPC-DS} benchmarks use the queries per hour and cost per performance metrics respectively.
    
    For benchmarking \textit{Graph} DMS, there are some existing works in their early stages (such as \textbf{HPC} Scalable Graph Analysis Benchmark~\cite{Dominguez-Sal:2010:SGD:1927585.1927590}, \textbf{Graph 500}~\cite{murphy2010introducing}, \textbf{XGDBench}~\cite{conf/cloudcom/DayarathnaS12}) dealing with 
    graph suitability transformations and graph analysis.
    However they fail to define standards for graph modeling and query languages.
    
    The substantial increase in the number of applications that use \textit{RDF} data has encouraged the need for large scale benchmarking efforts on all aspects of the Linked Data life cycle, mostly focusing on query processing~\cite{ngomo2016hobbit}.
    RDF DMS benchmarks make use of real (i.e., DBpedia or Wikidata) and synthetic (i.e., Berlin SPARQL Benchmark or WAT-DIV) datasets to evaluate DMS performance over custom stress-loads and setup environments \cite{fineeval}.\footnote{https://www.w3.org/wiki/RdfStoreBenchmarking}
    DBpedia SPARQL Benchmark \textbf{(DBPSB)}~\cite{Morsey2011} assesses RDF DMSs performance over DBpedia by creating a query workload derived from the DBpedia query logs. 
    
    The aim of the Lehigh University Benchmark (\textbf{LUBM})~\cite{Guo:2005:LBO:1741305.1741322} is to evaluate the performance of Semantic Web triple stores
    over a large synthetic dataset that complies to a university domain ontology.
    The Berlin SPARQL Benchmark (\textbf{BSBM}~\cite{Bizer2009TheBS}) is another benchmark based on synthetic data, which addresses e-commerce use cases built around a set of products offered by different vendors.
    The Waterloo SPARQL Diversity TEST Suite (\textbf{WatDiv}~\cite{alucc2014diversified}), provides data and query generators to enable benchmarking of RDF DMSs against a varying query structure (also complexity) to understand correlation of query typology with the variance in DMS performance.
    \textbf{SP2Bench}~\cite{books/sp/virgilio09/SchmidtHMPL09}, one of the most commonly used synthetic data based benchmarks,
    uses the schema of the DBLP bibliographic dataset\footnote{http://dblp.uni-trier.de/db/} to generate arbitrarily large datasets. 
   
    There are only a few efforts that benchmark \textit{cross-domain} DMS.
    \textbf{Pandora}\footnote{http://pandora.ldc.usb.ve/}, one such effort, uses the Berlin SPARQL Benchmark data to benchmark RDF stores against relational stores (Jena-TDB, Monetdb, GH-RDF-3X, PostgreSQL, 4Store). \textbf{Graphium}~\cite{flores2013graphium} is a similar study benchmarking RDF stores against Graph stores (Neo4J, Sparksee/DEX, HypergraphDB, RDF-3X) on graph datasets including a 10M triple graph data generated using the Berlin SPARQL Benchmark data generator.
    More recently, the \textbf{LDBC}~\cite{DBLP:journals/sigmod/AnglesBLF0ENMKT14} focuses on combining industry-strength benchmarks for graph and RDF data management systems.
    The LDBC introduces a new bottleneck methodology for developing benchmark workloads, which tries to combine user input with feedback from system experts.
    
    Research has so far focused on benchmarking domain specific DMSs, despite the need for integrating cross-domain DMSs and automating the benchmarking process.
    LITMUS aims at addressing these shortcomings and serving as an open, extensible platform to allow easy integration, benchmarking and performance analysis of diverse data management solutions.
    To the best of our knowledge, no such open, extensible and reusable framework exists, which allows to explore and analyse a wide range of different DMSs.

\section{Objectives, Challenges and Outcomes}\label{Objectives}
    \subsection{Focus of the LITMUS framework}
        The LITMUS framework aims at bridging the gaps in adopting, deploying and scaling the consumption of Linked Data. LITMUS focuses on simplifying the use, assessment and analysis of the performance of a wide spectrum of cross-domain DMSs. In particular, the LITMUS project will: 
        \begin{itemize}[nosep]
            \item \framebox[1.1\width]{\textbf{F1}} enable a common ground for benchmarking and comparing a plethora of cross-domain DMSs, and replicating existing third-party benchmarks;
            \item \framebox[1.1\width]{\textbf{F2}} create \textit{(i)} interoperable machine-readable evaluation reports and
            \textit{(ii)} scientific studies on the correlation of a variety of factors (such as query typology, data structures used for indexing, etc.) with respect to the performance of DMSs; 
            \item \framebox[1.1\width]{\textbf{F3}} recommend particular DMSs and benchmarks based on a set of user predefined requirements.  
        \end{itemize}

    \subsection{Challenges to be addressed}\label{challenges}
        To develop such an open extensible benchmarking platform, three key challenges have to be addressed:
        \begin{itemize}[nosep]
            \item \framebox[1.1\width]{\textbf{C1}} \textbf{Data conversion}: This challenge demands a generic data conversion mechanism allowing users to convert the RDF data to a format interpretable by the corresponding DMS.
            The focus is to represent RDF data in multiple formats, keeping the end user as secluded as possible from the framework's technical details. 
            \item \framebox[1.1\width]{\textbf{C2}} \textbf{Query Conversion:} 
            Cross-domain benchmarking of DMSs demands that queries are represented in all languages and formats supported by the respective tools.
            Query languages differ in their structure and expressivity. 
            For instance, complex path queries (in SPARQL, in particular Kleene stars) cannot be expressed in an equivalent SQL query.
            There is a need to develop an intermediate mechanism to convert or express the logic of one query (e.g. form SPARQL) to the other respective language (e.g. to CYPHER, SQL, CQL).
            This requires an exhaustive study of the query languages' specifications.
            The main challenge is to identify the correct mappings between different languages, maintaining the correctness and meaning of the original query.
            \item \framebox[1.1\width]{\textbf{C3}} \textbf{Performance indicators:} 
            The performance of a DMS can be assessed with regard to a variety of indicators.
            Dealing with the diverse characteristics of the DMSs, it is necessary to explore complex performance indicators in contrast to traditional ones, namely precision, recall, index size, storage size, number of triples, and query response time.
        \end{itemize}
    
    \subsection{Outcomes of LITMUS}
    \label{outcomes}
    The artifacts resulting from the LITMUS project will be \textit{(A1)} scientific studies and \textit{(A2)} frameworks/software. 
    
    \framebox[1.1\width]{\textbf{A1}}\textit{Scientific studies:} 
        \begin{itemize}[nosep]
            \item An in-depth analysis of the query language expressivity and supported features striving to address the language barrier \textbf{(C2)} (ref. section \ref{challenges}). This study will provide us with deep insights about the functionality of various query languages, their strengths and limitations.
            \item An exhaustive exploratory study on the selection of performance measures for evaluating cross-domain DMSs, addressing challenge\textbf{(C3)}(section \ref{challenges})
        \end{itemize}
        
        \framebox[1.1\width]{\textbf{A2}} \textit{Framework/Software (i.e. algorithms, tools, etc):} 
        \begin{itemize}[nosep]
            \item Automatic conversion of RDF data to multiple data formats (such as CSV, JSON, SQL, etc.), providing compatible data as input to the cross-domain DMSs. 
            \item Novel mechanisms for the automatic conversion of SPARQL to format-specific query languages, enabling compatible query input for cross-domain DMSs.
            \item An open, extensible benchmarking platform, LITMUS, for cross-domain DMS performance evaluation and easy replication of existing benchmarks.
       \end{itemize}       

    \subsection{Target audience}
            \textbf{Technology Vendors:}
            This addresses developers of commercial, industrial DMSs (including system and data analysts, system developers, system architects) who are thriving towards developing more and more advanced DMSs for efficient consumption of Big Data.\\
            \textbf{Technology Consumers:}
            Staff of private and commercial organisations and other users seeking recommendations for the best solution for their needs can simply compare a wide range of DMS against a list of desired parameters. \\
            \textbf{Technology Researchers:}
            The researchers who can benefit from our lessons-learned , and researchers whom LITMUS enables to contribute further results to the community.
            Target communities include Semantic Web, Databases, Information Retrieval, Big Data and others.

\section{The Litmus Framework}\label{litmus_framework}
      \subsection{Architecture Overview}
\begin{figure}[h]
            \centering
            \includegraphics[scale=0.14]{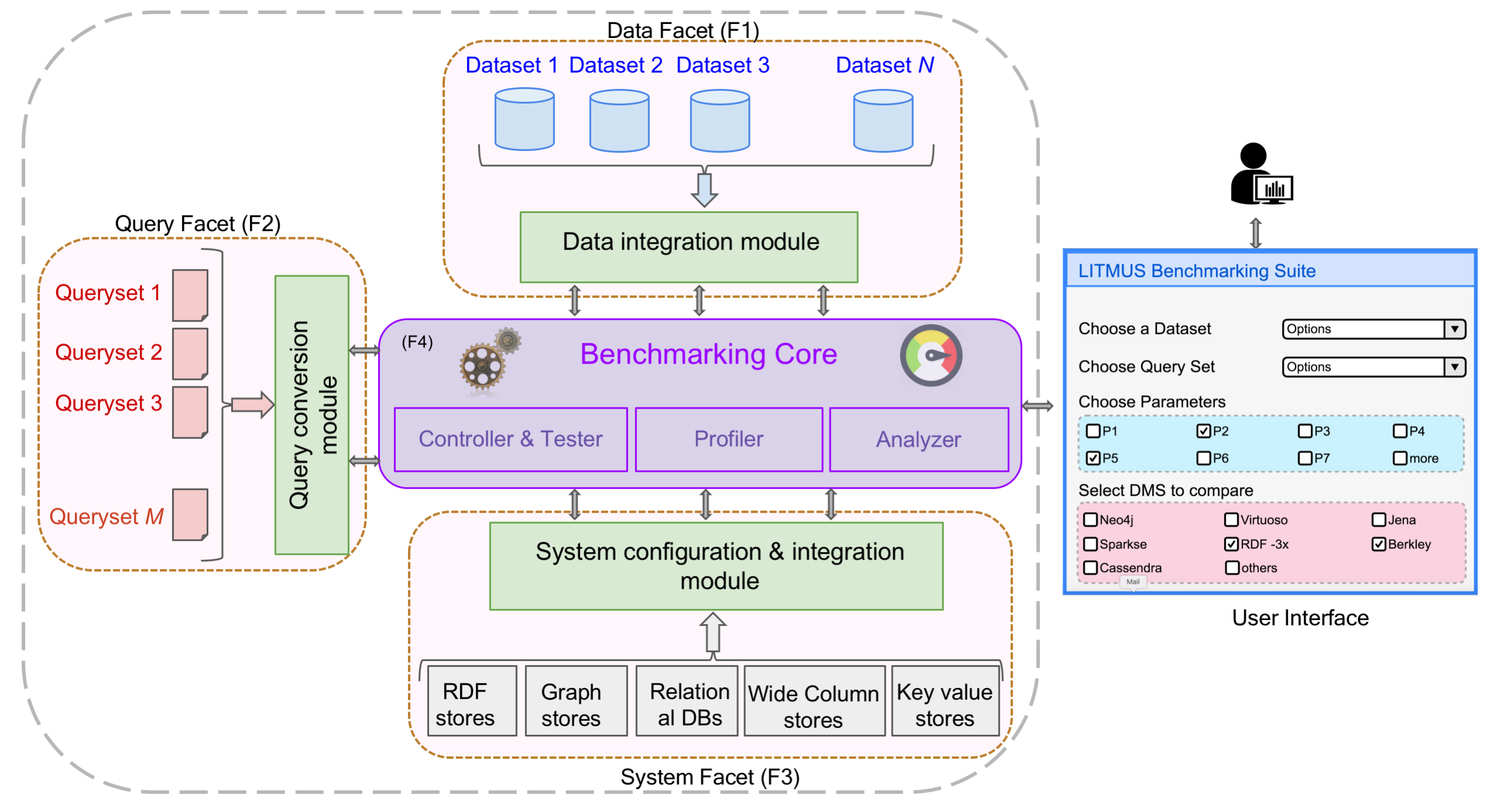}
            \caption{Overview of the LITMUS framework architecture.}
            \label{fig:benchmark_arch}
        \end{figure}

        The architecture of the LITMUS framework will comprise four major facets: Data Facet (F1), Query Facet (F2), System Facet (F3) and Benchmarking Core (F4) (Figure \ref{fig:benchmark_arch})
        In the following, we explain the role of each facet.
         
        \textbf{Data Facet \framebox[1.1\width]{F1}:} The Data Facet would deal with Dataset(s) and the Data Integration Module.
        Datasets chosen for benchmarking may be real datasets such as DBpedia\footnote{http://wiki.dbpedia.org/} Wikidata,\footnote{https://www.wikidata.org}; synthetic datasets such as the Berlin SPARQL Benchmarking (BSBM)~\cite{Bizer2009TheBS}, Waterloo SPARQL Diversity Test Suite (WatDiv)~\cite{alucc2014diversified}, or hybrid datasets comprising both real and synthetic data. 
        The \textit{Data Integration Module} is responsible for (a) making data available to the system in the requested formats (such as N-Triples, CSV, SQL or JSON) by carrying out appropriate data conversion and mapping tasks (cf. Challenge \textbf{C1}), and (b) loading the desired format of data to the respective DMSs selected for the benchmark. 
        
        \textbf{Query Facet \framebox[1.1\width]{F2}:} The Query Facet would deal with Queryset(s), and the Query Conversion Module. The \textit{Queryset} refers to the set of query input files. The \textit{Query Conversion Module} will be one of the key components addressing the language barrier (Challenge \textbf{C2}). It is responsible for converting the input SPARQL queries to the respective DMSs' query languages (such SQL, CYPHER or CQL).
        The conversion will be performed by developing an intermediate language/logic representation of the input query.
        The aim of this module is to allow efficient conversion of a wide variety of SPARQL queries (such as path, star-shaped and snowflake queries) to other query languages, ultimately breaking the language barrier.
        
        \textbf{System Facet \framebox[1.1\width]{F3}:} The System Facet also consists of two key modules, (i) DMSs and (ii) DMS Configuration and Integration module.
        The \textit{DMSs} module consists of the DMS selected for the benchmark.
        The \textit{DMS Configuration and Integration} module is responsible for (i) providing easy integration, via wrapper(s) or as a plugin, of the DMS, and (ii) monitor and configure the the integrated DMS for the benchmark.
        On top of this, this module will make use of \textit{Docker\footnote{Docker -- \url{https://www.docker.com/}} containers} to ensure a fair allocation of resources and to provide the necessary isolation required for conducting realistic benchmarks. 
        
        \textbf{Benchmarking Core \framebox[1.1\width]{F4}:} The Benchmarking Core is the heart of the LITMUS framework, consisting of three modules: (i) Controller and Tester, (ii) Profiler, and (iii) Analyser.
        The \textit{Controller and Tester} is responsible for executing the respective scripts for loading data and fetching the queries to their corresponding DMSs, creating and validating the specified system configurations, and finally executing the benchmark on the selected setting.
        The \textit{Profiler} is responsible for: (a) generating and loading various profiles (stress loads, query variations, etc.) for conducting the benchmark tests and (b) storing the benchmark results profile-wise. The \textit{Analyser} is responsible for collecting the benchmark results from the \textit{Profiler} and generates performance evaluation reports.
        It also performs a correlation analysis between the parameters specified by the user. The final results (reports) will then presented to the end user in a suitable visualisation.
        
\section{Conclusions}\label{conclusion} 
LITMUS addresses the gaps of the cross-benchmarking platform for different query languages and corresponding data management solutions. 
The literature review confirms the absence of such a cross-benchmarking platform.
We have mentioned the upcoming challenges, which the proposed system will have to address.
The proposed architecture of LITMUS would provide solutions to these challenges. \\
\textbf{Acknowledgements}: Parts of this work have been supported by the EU Horizon 2020 Framework Programme under grant agreements no. 642795 (WDAqua ITN), 644564 (Big Data Europe) and 688227 (HOBBIT).

\bibliographystyle{abbrv}
\bibliography{ref}

\begin{thebibliography}{10}

\bibitem{alucc2014diversified}
G.~Alu{\c{c}}, O.~Hartig, M.~T. {\"O}zsu, and K.~Daudjee.
\newblock Diversified stress testing of {RDF} data management systems.
\newblock In {\em International Semantic Web Conference}. Springer, 2014.

\bibitem{DBLP:journals/sigmod/AnglesBLF0ENMKT14}
R.~Angles, P.~A. Boncz, J.~Larriba{-}Pey, et~al.
\newblock The linked data benchmark council: a graph and {RDF} industry
  benchmarking effort.
\newblock {\em {SIGMOD} Record}, 2014.

\bibitem{Bizer2009TheBS}
C.~Bizer and A.~Schultz.
\newblock The berlin {SPARQL} benchmark.
\newblock {\em Int. J. Semantic Web Inf. Syst.}, 5, 2009.

\bibitem{conf/cloudcom/DayarathnaS12}
M.~Dayarathna and T.~Suzumura.
\newblock {XGDBench}: A benchmarking platform for graph stores in exascale
  clouds.
\newblock In {\em CloudCom}. IEEE Computer Society, 2012.

\bibitem{Dominguez-Sal:2010:SGD:1927585.1927590}
D.~Dominguez-Sal, P.~Urb\'{o}n-Bayes, A.~Gim{\'e}nez-Va\~{n}\'{o}, et~al.
\newblock Survey of graph database performance on the {HPC} scalable graph
  analysis benchmark.
\newblock In {\em Proceedings of the 2010 International Conference on Web-age
  Information Management}, WAIM'10. Springer-Verlag, 2010.

\bibitem{flores2013graphium}
A.~Flores, G.~Palma, M.-E. Vidal, et~al.
\newblock {GRAPHIUM}: visualizing performance of graph and rdf engines on
  linked data.
\newblock In {\em Proceedings of the 2013th International Conference on Posters
  \& Demonstrations Track-Volume 1035}. CEUR-WS. org, 2013.

\bibitem{Guo:2005:LBO:1741305.1741322}
Y.~Guo, Z.~Pan, and J.~Heflin.
\newblock {LUBM}: A benchmark for owl knowledge base systems.
\newblock {\em Web Semant.}, 3, Oct. 2005.

\bibitem{Morsey2011}
M.~Morsey, J.~Lehmann, S.~Auer, and A.-C. Ngonga~Ngomo.
\newblock {\em DBpedia {SPARQL} Benchmark -- Performance Assessment with Real
  Queries on Real Data}.
\newblock Springer Berlin Heidelberg, 2011.

\bibitem{murphy2010introducing}
R.~C. Murphy, K.~B. Wheeler, B.~W. Barrett, and J.~A. Ang.
\newblock Introducing the {GRAPH} 500.
\newblock {\em Cray User’s Group (CUG)}, 2010.

\bibitem{Nambiar2011}
R.~Nambiar, N.~Wakou, F.~Carman, and M.~Majdalany.
\newblock {\em Transaction Processing Performance Council {(TPC)}: State of the
  Council 2010}.
\newblock Springer Berlin Heidelberg, 2011.

\bibitem{ngomo2016hobbit}
A.-C.~N. Ngomo and M.~R{\"o}der.
\newblock {HOBBIT}: Holistic benchmarking for big linked data.
\newblock {\em ERCIM News}, 2016.

\bibitem{fineeval}
M.~Saleem, Y.~Khan, A.~Hasnain, I.~Ermilov, and A.~N. Ngomo.
\newblock A fine-grained evaluation of {SPARQL} endpoint federation systems.
\newblock {\em Semantic Web}, 7, 2015.

\bibitem{books/sp/virgilio09/SchmidtHMPL09}
M.~Schmidt, T.~Hornung, M.~Meier, C.~Pinkel, and G.~Lausen.
\newblock {SP2Bench}: A {SPARQL} performance benchmark.
\newblock In {\em Semantic Web Information Management}. Springer, 2009.

\bibitem{bioasq}
G.~Tsatsaronis, G.~Balikas, P.~Malakasiotis, et~al.
\newblock An overview of the {BIOASQ} large-scale biomedical semantic indexing
  and question answering competition.
\newblock {\em {BMC} Bioinformatics}, 16, 2015.

\bibitem{qald}
C.~Unger, C.~Forascu, V.~Lopez, et~al.
\newblock Question answering over linked data {(QALD-5)}.
\newblock In {\em Working Notes of {CLEF} 2015, Toulouse, France, September
  8-11}, 2015.

\bibitem{gerbil}
R.~Usbeck, M.~R{\"{o}}der, A.~N. Ngomo, et~al.
\newblock {GERBIL:} general entity annotator benchmarking framework.
\newblock In {\em Proceedings of the 24th International Conference on World
  Wide Web, {WWW} 2015, Florence, Italy, May 18-22}, 2015.

\end{thebibliography}

\end{document}